\documentclass{article}

\usepackage{amsmath}
\usepackage{amsthm}
\usepackage{amssymb}
\usepackage{bm}
\usepackage{array}
\usepackage{cite}
\usepackage{csquotes}
\usepackage{soul}
\usepackage{authblk}

\begin{document}

\title{Stochastic differential equation model for spontaneous emission and carrier noise in semiconductor lasers}

\author{Austin McDaniel\thanks{Austin.McDaniel@asu.edu} \hspace{1pt}}
        
\author{Alex Mahalov\thanks{mahalov@asu.edu}}

\affil{School of Mathematical and Statistical Sciences, Arizona State University, Tempe, AZ 
              85287, USA}

\date{\vspace{-5ex}}

\maketitle

\begin{abstract}
We present a new stochastic differential equation model for the spontaneous emission noise and carrier noise in semiconductor lasers.  The correlations between these two types of noise have often been neglected in 
recent studies of the effects of the noise on the laser dynamics.  However, the classic results of Henry show that the intensity noise and the carrier noise 
are strongly negatively correlated.  Our model demonstrates
how to properly account for these correlations since the 
corresponding diffusion coefficients agree exactly with those derived by Henry.  We show that in fact in the correct model the spontaneous emission noise and the carrier noise are driven by the same Wiener 
processes.  Furthermore, we demonstrate that the nonzero correlation time of the physical noise affects the mean dynamics of both the electric field amplitude and 
the carrier number.  We show that these are systematic corrections that can be described by additional drift terms in the model.
\end{abstract}

\section{Introduction}

The inherent noise in semiconductor laser systems is a complex topic \cite{Scully1966, Scully1967, Lax1966, Lax, Henry1982, Henry1983, Chow, Carmichael, Meystre} that has received 
renewed attention due to recent interest in the 
dynamics of semiconductor lasers with optical feedback \cite{Mulet, Giacomelli, Buldu2001, Buldu2002, Buldu2004, Marino, Yousefi, Torcini, Soriano2011}. 
Spontaneous recombination within the gain medium leads to two types 
of intrinsic noise: spontaneous emission noise and carrier noise.  The former, also referred to as field noise, results from emitted photons in the 
lasing mode, and the latter, also referred to as shot noise, is due to the discrete and instantaneous nature of these events \cite{Yousefi, VanderSande}.  Spontaneous emission noise 
affects the linewidth of a semiconductor laser \cite{Henry1982, Henry1983, Scully1967} and can give rise to the phenomenon of mode hopping in which the laser transitions from, e.g., 
multi-mode to single-mode operation \cite{Yamada}.  Enhanced spontaneous emission and linewidth are distinguishing features of semiconductor lasers compared to other types of lasers and, in particular, 
contribute to the extreme sensitivity of semiconductor lasers to optical feedback \cite{Kane}.  

In this paper we consider the problem of modeling the intrinsic noise in semiconductor lasers within the framework of stochastic differential equations (SDEs).  Spontaneous emission noise 
is often accounted for in models of the semiconductor laser system by including a Langevin noise term in the field equation \cite{Mulet, Buldu2001, Buldu2002, Buldu2004, Torcini}.  Such a system
that contains Langevin noise terms is rigorously interpreted by defining Wiener processes that 
drive the noise.  The advantage of introducing this additional mathematical framework is that 
the rich theory of stochastic differential equations (see, e.g., \cite{Gardiner, Oksendal}) can then be employed to analyze the physical system.  In Ref. \cite{Carmichael} a Fokker-Planck equation for the 
field amplitude was derived, and from this the corresponding SDE accounting for the spontaneous emission noise was obtained.  In addition to
spontaneous emission noise, carrier noise has also been considered within the dynamical model by the inclusion of another 
Langevin noise term in the equations (see, e.g., \cite{Yousefi, Soriano2011}).  However, an SDE model 
that precisely describes the relationship between the Wiener processes that drive both types of noise has not been derived or presented.  We show in this paper that the SDE framework is necessary in order to correctly
account for the correlations between the two types of noise within the dynamical model of the semiconductor laser system.

In numerical studies of the effects of the intrinsic noise on the laser dynamics, typically the correlations between the spontaneous emission noise
and the carrier noise either have not been considered or have been explicitly neglected under the presumption that they are negligible.  However, the classic work of Henry shows that these two types of noise are 
closely related.  In Refs. \cite{Henry1982, Henry1983}, Henry derived the steady-state diffusion coefficients associated with the intensity and phase of the field and the carrier number.  These results show 
that while the phase noise and the carrier noise are uncorrelated, the intensity noise and the carrier noise are strongly negatively correlated.  While these diffusion coefficients characterize the 
statistical distribution of the noise, they do not describe the effects of the noise on the laser dynamics, for which an SDE model is necessary.  In this paper we present a new SDE model for which the
corresponding diffusion coefficients agree exactly with those derived by Henry.  We therefore show how to correctly account for, within the dynamical model, the strong negative correlation between the intensity 
noise and the carrier noise.  In particular, in this correct model the same Wiener processes that drive the spontaneous emission noise also drive the carrier noise.

We begin with the rate equations for the complex amplitude $E$ of the electric field inside the cavity and 
the carrier number $N$, in which we include terms representing the spontaneous emission noise and the carrier noise.  We consider the more general model that includes a delayed feedback term that
represents optical feedback.  Our starting point is therefore the system (see, e.g., \cite{Mulet, Buldu2001, Buldu2002, Buldu2004})
\begin{subequations}
\label{LKequations}
\begin{align}
\label{electric field}
 \frac{ d E }{dt} (t) = \; & \frac{ 1 + i \alpha}{2} \Big( G \big( E(t) , N(t) \big) - \gamma \Big) E (t) \notag \hspace{60pt} \\
     & \hspace{55pt} + \kappa e ^{- i \omega \delta } E (t - \delta ) + \zeta (t)  \\
 \label{carrier number}
 \frac{ d N }{dt} (t) = \; & \gamma _e \Big( C N ^{\mathrm{th}} - N(t) \Big) \notag \\
                      & \hspace{28pt} - G \big( E(t) , N(t) \big) \left| E (t) \right| ^2 + \xi (t)
\end{align}
\end{subequations}
where $G$ is the material gain function given by
\begin{equation}
\label{gain function}
  G \big( E(t) , N(t) \big) = \frac{ g \big[ N(t) - N^0 \big] }{1 + s | E(t) | ^2 } \; . \hspace{15pt}
\end{equation}
In (\ref{LKequations}), $\alpha$ is the linewidth enhancement factor, $\gamma$ is the inverse photon lifetime, 
$\kappa$ is the feedback strength, $\delta$ is the feedback roundtrip time, $\omega$ is the emitting 
frequency of the laser without feedback, $\gamma _e$ is the inverse carrier lifetime, $C$ is the pump current, and $ N ^{\mathrm{th}}$ is the 
threshold carrier number.  Note that the system without feedback can be obtained by simply setting $\kappa = 0$.  The last term in (\ref{electric field}) and that in (\ref{carrier number}) represent 
the spontaneous emission noise and carrier noise, respectively, and are the focus of this paper. In (\ref{gain function}), $g$ is the differential gain coefficient, $N^0$ is the 
carrier number at transparency, and $s$ is the saturation coefficient \cite{Buldu2004}.  

\section{Stochastic differential equation model}

When spontaneous emission noise and carrier noise are included in the model, the noise terms $\zeta$ and $\xi$ are typically taken to be mean-zero
Gaussian white noises \cite{Mulet, Buldu2001, Buldu2002, Buldu2004, Soriano2011, Yousefi}, with $\zeta$ being complex-valued \cite{Torcini}.  The spontaneous emission noise $\zeta$ depends on   
the spontaneous emission rate $R = 4 \beta N$ \cite{Barland, Mulet} where the parameter $\beta$ controls the strength of the noise.  In \cite{Carmichael} it was shown
that the Wiener process that drives the real part of the 
spontaneous emission noise $\zeta$ is independent of the (real-valued) Wiener process that drives the imaginary part of $\zeta$, and this has been taken into account in studies of 
the effects of the noise on the semiconductor laser dynamics (see, e.g., \cite{Torcini}).  We will show that this result is consistent with those of Henry, which we now discuss.  Let the complex 
amplitude be expressed as $E_t = | E_t | e ^{i \theta _t}$ where the phase $\theta _t$ is real and let $I_t = | E_t | ^2$ denote the intensity of the electric field.  Henry derived the steady-state diffusion 
coefficients corresponding to the intensity, phase, and carrier number, which are expressed as follows in terms of the 
intensity $I$, the spontaneous emission rate $R$, and the electron-hole recombination rate $S$ \cite{Henry1982, Henry1983}:
\begin{align}
\label{diffusion coefficients}
 & D_{II} = R I , \hspace{15pt} D_{\theta \theta} = \frac{R}{4I} \hspace{1pt} , \hspace{15pt} D_{NN} = R I + S \hspace{1pt} , \notag \\
 & \hspace{23pt} D_{IN} = - R I , \hspace{20pt}  D_{I \theta} = D_{\theta N} = 0 \; .
\end{align}
We will discuss below the precise meaning of the diffusion coefficients given by (\ref{diffusion coefficients}).  First, we introduce an SDE model for which the
corresponding diffusion coefficients agree exactly with (\ref{diffusion coefficients}).
Let $E^R$ and $E^I$ denote the real and imaginary parts, respectively, of $E$.  In order to state our model and the results of the next section as generally as possible we let
the electron-hole recombination rate $S$ be an arbitrary function of the carrier number $N$.  We will show that the diffusion coefficients given by (\ref{diffusion coefficients})
correspond to the noise terms in the following SDE system: 
\begin{subequations}
\label{LK SDE}
\begin{align}
\label{electric field SDE}
  d E _t = & \; \frac{ 1 + i \alpha}{2} \Big( G \big( E_t , N_t \big) - \gamma \Big) E _t dt + \kappa e ^{- i \omega \delta } E _{t - \delta } \hspace{1pt} dt \notag \hspace{20pt} \\
           & \hspace{60pt} + \sqrt{2 \beta N_t } \Big( dW^A _t + i dW^B _t \Big) \\
 d N _ t  = & \; \gamma _e \Big( C N ^{\mathrm{th}} - N_t \Big) dt  - G \big( E_t , N_t \big) \left| E _t \right| ^2 dt \notag \hspace{20pt} \\
              & - 2 \sqrt{2 \beta N_t } \Big( E^R _t dW^A _t + E^I _t dW^B _t \Big) \notag \\
              & \hspace{100pt} + \sqrt{2S (N_t)} dW^C_t
\end{align}
\end{subequations}
where the Wiener processes $W^A , W^B ,$ and $W^C$ are mutually independent (here and in the rest of the paper we use standard SDE notation).  Note that the noise terms in both equations are multiplicative since the
spontaneous emission noise depends on the carrier number through the spontaneous emission rate and the carrier noise depends on both the field and the carrier number.
We also note that in the presence of optical feedback ($\kappa \neq 0$), (\ref{LK SDE}) is a stochastic differential delay equation (SDDE) system.  
Therefore, in this case, its solution is non-Markovian.

An SDE model for the intensity and the phase can be found from (\ref{electric field SDE}).
The equation for the intensity is found by using the It\^o product formula:
\begin{equation*}
dI _t = d \left( E_t \overline{E_t} \right) = E_t d \overline{E_t} + \overline{E_t} d E_t + \big( d E_t \big) \big( d \overline{E_t} \big)
\end{equation*}
where the last term is computed using the It\^o calculus \cite{Gardiner, Oksendal}.  The equation for the phase can then be obtained by using the It\^o formula.  The resulting equations for 
the intensity and the phase are
\begin{subequations}
\label{intensity and phase}
\begin{align}
 \label{intensity equation}
 d I_t = \; &  \Big( G \big( I_t , N_t \big) - \gamma \Big) I_t dt 
                      + 2 \kappa \mathrm{Re} \Big( e ^{ i \omega \delta } E_t \overline{ E _{t - \delta } } \Big) dt \notag \hspace{20pt} \\
                   & + 4 \beta N_t dt + 2 \sqrt{2 \beta N_t } \Big( E^R _t dW ^A _t + E^I _t dW ^B _t \Big) \\           
 \label{phase equation}
 d \theta _t = \; &   \frac{ \alpha}{2} \Big( G \big( I_t , N_t \big) - \gamma \Big) dt 
                   - \kappa \mathrm{Re} \left( i e ^{- i \omega \delta } \frac{E _{t - \delta }}{ E_t } \right) dt \notag \hspace{20pt} \\
                   & \hspace{45pt} + \frac{\sqrt{2 \beta N_t }}{I_t} \Big( E^R _t dW^B _t - E^I _t dW^A _t \Big)
\end{align}
\end{subequations}
where $\mathrm{Re} (z)$ denotes the real part of the complex number $z$, and where we have explicitly denoted the fact that the material gain function $G$ depends on $E_t$ only
through $I_t$.  

We discuss here the case $\kappa = 0$ in which the solution of (\ref{LK SDE}) is a Markov process so that there is an associated Fokker-Planck equation that describes the 
evolution of the probability density function of the system's state \cite{Gardiner} (there is no 
Fokker-Planck equation for the non-Markovian system corresponding to the case $\kappa \neq 0$).  In 
the white-noise driven model (\ref{LK SDE}) the noise terms give rise to diffusion terms in the corresponding Fokker-Planck 
equation.  The coefficients of these diffusion terms can be found from the stochastic differential equation system as follows.  In the Fokker-Planck equation, $E$ and $N$ are independent variables, and so we 
let $\big< \cdot \big>$ denote the expectation operator that treats these variables as such, i.e., as deterministic quantities.
Let $F_I, F_{\theta},$ and $F_N$ denote the intensity, phase, and carrier noises, respectively, that is, 
\begin{subequations}
\label{noise terms}
\begin{flalign}
F_I (t) dt & = 2 \sqrt{2 \beta N_t } \Big( E^R _t dW ^A _t + E^I _t dW ^B _t \Big) \; , \hspace{10pt}  \\
F_{\theta} (t) dt & = \frac{\sqrt{2 \beta N_t }}{I_t} \Big( E^R _t dW^B _t - E^I _t dW^A _t \Big) \; , \\
\mathrm{and} \hspace{30pt} & \notag \\
F_N (t) dt & = - 2 \sqrt{2 \beta N_t } \Big( E^R _t dW^A _t + E^I _t dW^B _t \Big) \notag \hspace{50pt} \\
              & \hspace{90pt} + \sqrt{2S (N_t)} dW^C_t \; .
\end{flalign}
\end{subequations}
For $j, k = I, \theta, N$, the 
diffusion coefficients $D_{jk}$ are then given by
\begin{equation}
\label{definition diffusion coefficients}
 \big< F_j (t) F_k (s) \big> = 2 D_{jk} \delta (t - s) \; .
\end{equation}
The diffusion coefficients corresponding to the SDE system can be found by using (\ref{noise terms}), (\ref{definition diffusion coefficients}), and the independence of the Wiener processes $W^A , W^B ,$ and $W^C$.
The diffusion coefficients obtained this way agree exactly with those derived by Henry that are given in (\ref{diffusion coefficients}).  

\section{Systematic corrections due to nonzero correlation time}

In the system (\ref{LK SDE}) the spontaneous emission noise and the carrier noise are represented by white noises, i.e., their correlation functions are taken to be 
delta functions so that their 
correlation times are equal to zero.  Physically, the Langevin reservoir forces are not, however, delta-correlated; their correlation time is nonzero.  Nonetheless, since their 
correlation time is small compared to all of the typical relaxation times of the system \cite{Lax1966} they are normally modeled by white noise.  However, when the noise in a system 
is multiplicative (state-dependent), the nonzero correlation time of the physical noise can affect the mean dynamics of the system.  Furthermore, the magnitude of these effects does not depend on the 
size of the small correlation time, that is, these effects do not vanish for exceedingly small correlation times.  Mathematically, these effects can be seen by using a colored noise 
process (i.e., a process having a nonzero correlation 
time) to model the physical noise and subsequently taking the white-noise limit of the system.  This procedure gives rise to additional drift terms in the system called the Stratonovich corrections
that capture the effects of the nonzero correlation time on the mean dynamics.  This fact is sometimes referred to as the Wong-Zakai theorem \cite{Wong1965} 
and is discussed in, e.g., Gardiner's text \cite{Gardiner}. 

Because the spontaneous emission noise and the carrier noise are multiplicative, such effects are present in the semiconductor laser system.  The purpose of this section is to derive these corrections.  
To do this we replace the white noises in 
(\ref{LK SDE}) with noise processes having small but nonzero correlation times.  
As a model for the noise 
we use the stationary Ornstein-Uhlenbeck colored noise process.  This choice is a good model of the physical noise since it is 
a Gaussian process with a rapidly (exponentially) decaying correlation function.  For $i = A, B, C$, we define 
the process $\eta ^i$ as the stationary solution of the SDE
\begin{equation}
\label{OU process}
 d \eta ^i _t = - \frac{1}{\tau} \eta ^i _t dt + \frac{1}{\tau} dW ^i _t \hspace{10pt}
\end{equation}
where $\tau > 0$.  The stationary solution of (\ref{OU process}) is the solution corresponding to the initial condition that has a mean-zero Gaussian distribution with 
variance $ (2 \tau )^{-1}$.  Defined this way, each $\eta ^i$ is a mean-zero Gaussian process with
autocovariance function 
\begin{equation}
\label{covariance function}
	E \big[ \eta ^i _t \eta ^i _s \big] = K ( t - s )
	= \frac{1}{2 \tau } e^{- \frac{|t-s|}{\tau}}  \hspace{10pt}
\end{equation} 
(we let $E[ \cdot ]$ denote expectation, in contrast to the electric field amplitude $E$ which is never followed by brackets in this paper).  For $i \neq j$, the 
processes $\eta ^i$ and $\eta ^j$ are independent by the independence of $W^i$ and $W^j$.  In view of (\ref{covariance function}) each process $\eta ^i$ has correlation 
time $\tau$, and $K( t - s) \rightarrow \delta (t - s)$ as $\tau \rightarrow 0$.  More precisely, 
as $\tau \rightarrow 0$,
\begin{equation*}
 \int _0 ^t \eta ^i _s ds \hspace{3pt} \longrightarrow \hspace{3pt} W^i _t 
\end{equation*}
(see, e.g., \cite{Nelson}).  This justifies replacing
in (\ref{LK SDE}) each $dW^i _t$ with $\eta ^i _t dt$ to get
\begin{subequations}
\label{SDE2}
\begin{align}
\label{electric field SDE2}
  d E _t = & \; \frac{ 1 + i \alpha}{2} \Big( G \big( E_t , N_t \big) - \gamma \Big) E _t dt \notag \hspace{90pt} \\
                   & + \kappa e ^{- i \omega \delta } E _{t - \delta } \hspace{1pt} dt
                        + \sqrt{ 2 \beta N_t } \Big( \eta ^A _t + i \eta ^B _t \Big) dt  \\
\label{carrier number SDE2}
 d N _ t = & \; \gamma _e \Big( C N ^{\mathrm{th}} - N_t \Big) dt - G \big( E_t , N_t \big) \left| E _t \right| ^2 dt \notag \\
           &  - 2 \sqrt{2 \beta N_t } \Big( E^R _t \eta ^A _t + E^I _t \eta ^B _t \Big) dt \notag \\
           & \hspace{100pt} + \sqrt{2S( N_t )} \eta ^C_t dt \; .
\end{align}
\end{subequations}

We now derive the limit of the system (\ref{SDE2}) as the correlation time $\tau$ of the noises goes to zero.
We show that the resulting white-noise driven system is
\begin{subequations}
\label{limiting system}
\begin{align}
\label{limiting equation 1}
  d E _t = & \; \frac{ 1 + i \alpha}{2} \Big( G \big( E_t , N_t \big) - \gamma \Big) E _t dt + \kappa e ^{- i \omega \delta } E _{t - \delta } \hspace{1pt} dt \notag \hspace{20pt} \\
           & \hspace{20pt} - \beta E_t dt + \sqrt{2 \beta N_t } \Big( dW^A _t + i dW^B _t \Big) \\
\label{limiting equation 2}
 d N _ t  = & \; \gamma _e \Big( C N ^{\mathrm{th}} - N_t \Big) dt  - G \big( E_t , N_t \big) \left| E _t \right| ^2 dt \notag \hspace{20pt} \\
              & + 2 \beta \left| E _t \right| ^2 dt - 4 \beta N_t dt + \frac{1}{2} S'(N_t) dt \notag \\
              & - 2 \sqrt{2 \beta N_t } \Big( E^R _t dW^A _t + E^I _t dW^B _t \Big) \notag \\
              & \hspace{100pt} + \sqrt{2S (N_t)} dW^C_t 
\end{align}
\end{subequations}
where $S'$ is the derivative of the electron-hole recombination rate.  Note that the system (\ref{limiting system}) is not equal to (\ref{LK SDE}); there are additional drift terms in 
(\ref{limiting system}) that are due to the dependence of the strengths of the noises on the carrier number and the field.
These terms are $- \beta E_t dt$ in (\ref{limiting equation 1}) and $ 2 \beta \left| E _t \right| ^2 dt , - 4 \beta N_t dt ,$ and $\frac{1}{2} S'(N_t) dt$ in 
(\ref{limiting equation 2}).  
We now turn to the derivation of (\ref{limiting system}).

\subsection{Derivation}

For convenience we let
 $f \left( E_t , E_{t - \delta} , N_t \right) dt$ denote the first two terms on the right-hand side of (\ref{electric field SDE2}) and
 $h \left( E_t , N_t \right) dt$ denote the first two terms on the right-hand side of (\ref{carrier number SDE2}).  We solve (\ref{OU process}) for $\eta ^i _t dt$ and 
 substitute the resulting expression into (\ref{SDE2}) to get 
\begin{align*}
  d E _t = & \; f \big( E_t , E_{t - \delta} , N_t \big) dt - \sqrt{ 2 \beta N_t } \Big( \tau d \eta ^A _t + i \tau d \eta ^B _t \Big) \notag \hspace{15pt} \\
            &    \hspace{95pt} + \sqrt{ 2 \beta N_t } \Big( dW ^A _t + i dW ^B _t \Big) \\
 d N _ t = & \; h \big( E_t , N_t \big) dt + 2 \sqrt{ 2 \beta N_t } \Big( E^R_t \tau d \eta ^A _t + E^I_t \tau d \eta ^B _t \Big) \\
              & - 2 \sqrt{ 2 \beta N_t } \Big( E^R_t dW ^A _t + E^I _t dW^B _t \Big) \\
              & \hspace{55pt} - \sqrt{2S (N_t) } \tau d \eta ^C _t + \sqrt{2S(N_t)} dW^C_t \; .
\end{align*}
It is useful to work with the integral form of this system:
\begin{subequations}
\begin{flalign}
 \label{integral equation 1}
  E _t = & \; E_0 + \int_0 ^t f \big( E_s , E_{s - \delta} , N_s \big) ds \notag \hspace{100pt} \\
     & - \int_0 ^t \sqrt{ 2 \beta N_s } \Big( \tau d \eta ^A _s + i \tau d \eta ^B _s \Big) \notag \\
            &    \hspace{40pt} + \int_0 ^t \sqrt{ 2 \beta N_s } \Big( dW ^A _s + i dW ^B _s \Big) \\
 \label{integral equation 2}
 N _ t = & \; N_0 + \int_0 ^t h \big( E_s , N_s \big) ds \notag \\
            & + \int_0 ^t 2 \sqrt{ 2 \beta N_s } \Big( E^R_s \tau d \eta ^A _s + E^I_s \tau d \eta ^B _s \Big) \notag \\
             &  - \int_0 ^t 2 \sqrt{ 2 \beta N_s } \Big( E^R_s dW ^A _s + E^I _s dW^B _s \Big) \notag \\ 
             & - \int_0 ^t \sqrt{2S (N_s)} \tau d \eta ^C _s +  \int_0 ^t \sqrt{2S (N_s)} dW^C_s \; . 
\end{flalign}
\end{subequations}
We integrate by parts the second integral in (\ref{integral equation 1}) and the second and fourth integrals in 
(\ref{integral equation 2}) to get
\begin{subequations}
\label{after by parts}
\begin{flalign}
  E _t = & \; E_0  - \sqrt{ 2 \beta N_t } \Big( \tau \eta ^A _t + i \tau \eta ^B _t \Big) \notag \hspace{100pt} \\ 
     & + \sqrt{ 2 \beta N_0 } \Big( \tau \eta ^A _0 + i \tau \eta ^B _0 \Big) 
                + \int_0 ^t f \big( E_s , E_{s - \delta} , N_s \big) ds \notag \\
     & + \int_0 ^t \sqrt{ \frac{\beta}{ 2 N_s }} \Big( \tau \eta ^A _s + i \tau \eta ^B _s \Big) d N _s \notag \\
            &    \hspace{50pt} + \int_0 ^t \sqrt{ 2 \beta N_s } \Big( dW ^A _s + i dW ^B _s \Big) \\[2pt]
  N _ t = & \; N_0 + 2 \sqrt{ 2 \beta N_t } \Big( E^R_t \tau \eta ^A _t + E^I_t \tau \eta ^B _t \Big) \notag \\ 
                  & - 2 \sqrt{ 2 \beta N_0 } \Big( E^R_0 \tau \eta ^A _0 + E^I_0 \tau \eta ^B _0 \Big)
                   - \sqrt{2S (N_t)} \tau \eta ^C _t \notag \\ 
                   & + \sqrt{2S (N_0)} \tau \eta ^C _0 + \int_0 ^t h \big( E_s , N_s \big) ds \notag \\
            & - \int_0 ^t \sqrt{ \frac{ 2 \beta }{ N_s }} 
                                  \Big( E^R_s \tau \eta ^A _s + E^I_s \tau \eta ^B _s \Big) d N_s \notag \\
            & - \int_0 ^t 2 \sqrt{ 2 \beta N_s } \Big(  \tau \eta ^A _s dE^R_s + \tau \eta ^B _s dE^I_s \Big) \notag \\
             &  - \int_0 ^t 2 \sqrt{ 2 \beta N_s } \Big( E^R_s dW ^A _s + E^I _s dW^B _s \Big) \notag \\ 
             & + \int_0 ^t \frac{S'(N_s)}{\sqrt{2S (N_s)}} \tau \eta ^C _s dN_s
                 +  \int_0 ^t \sqrt{2S (N_s)} dW^C_s \; . 
\end{flalign}
\end{subequations}
Note that since the solution $(E, N)$ of (\ref{SDE2}) is differentiable the It\^o terms arising from the stochastic integration by parts formula are zero.
At this point we need the equations for $E^R$ and $E^I$ which follow from (\ref{electric field SDE2}):
\begin{subequations}
\label{real and imaginary field}
\begin{align}
\label{real field}
  d E ^R _t = & \hspace{4pt} \mathrm{Re} \Big( f \big( E_t , E_{t - \delta} , N_t \big) \Big) dt 
                       + \sqrt{ 2 \beta N_t } \eta ^A _t dt  \hspace{10pt} \\
 \label{imaginary field}
d E ^I _t = & \hspace{4pt} \mathrm{Im} \Big( f \big( E_t , E_{t - \delta} , N_t \big) \Big) dt 
                   + \sqrt{ 2 \beta N_t } \eta ^B _t dt \; .
\end{align}
\end{subequations}
In (\ref{after by parts}) we substitute (\ref{carrier number SDE2}), (\ref{real field}), and (\ref{imaginary field}) for 
$d N _s, d E^R_s,$ and $dE ^I_s$ to get
\begin{align*}
  E _t = & \; E_0  + U^E_t + \int_0 ^t f \big( E_s , E_{s - \delta} , N_s \big) ds \notag \\
     & + H^E _t + \int_0 ^t \sqrt{ 2 \beta N_s } \Big( dW ^A _s + i dW ^B _s \Big) \hspace{30pt} \\
  N _ t = & \; N_0 + U^N _t + \int_0 ^t h \big( E_s , N_s \big) ds + H^N _t \notag \\
             &  - \int_0 ^t 2 \sqrt{ 2 \beta N_s } \Big( E^R_s dW ^A _s + E^I _s dW^B _s \Big) \notag \\ 
             &  +  \int_0 ^t \sqrt{2S (N_s)} dW^C_s
\end{align*}
where
\begin{align*}
U^E _t = & \hspace{1pt} - \sqrt{ 2 \beta N_t } \Big( \tau \eta ^A _t + i \tau \eta ^B _t \Big) 
            + \sqrt{ 2 \beta N_0 } \Big( \tau \eta ^A _0 + i \tau \eta ^B _0 \Big) \notag \\
  & + \int_0 ^t \sqrt{ \frac{\beta}{ 2 N_s }} h \big( E_s , N_s \big) \Big( \tau \eta ^A _s + i \tau \eta ^B _s \Big) ds \; ,
\end{align*}
\begin{align*}
H^E _t = & \hspace{1pt} - \int_0 ^t 2 \beta \Big( \tau \eta ^A _s + i \tau \eta ^B _s \Big) 
                    \Big( E^R _s \eta ^A _s + E^I _s \eta ^B _s \Big) ds \hspace{30pt} \\
             & + \int_0 ^t \sqrt{ \frac{ \beta S( N_s ) }{ N_s }} 
                         \Big( \tau \eta ^A _s + i \tau \eta ^B _s \Big) \eta ^C_s ds \; ,
\end{align*}
\begin{align*}
 U^N _t = & \hspace{4pt} 2 \sqrt{ 2 \beta N_t } \Big( E^R_t \tau \eta ^A _t + E^I_t \tau \eta ^B _t \Big) \hspace{90pt} \\ 
                  & - 2 \sqrt{ 2 \beta N_0 } \Big( E^R_0 \tau \eta ^A _0 + E^I_0 \tau \eta ^B _0 \Big) \\
                   & - \sqrt{2S (N_t)} \tau \eta ^C _t + \sqrt{2S (N_0)} \tau \eta ^C _0 \\
                   & - \int_0 ^t \sqrt{ \frac{ 2 \beta }{ N_s }} \hspace{1pt} h \big( E_s , N_s \big)
                                  \Big( E^R_s \tau \eta ^A _s + E^I_s \tau \eta ^B _s \Big) ds \\
                  & - \int_0 ^t 2 \sqrt{ 2 \beta N_s } \hspace{1pt} 
                                \mathrm{Re} \Big( f \big( E_s , E_{s - \delta} , N_s \big) \Big) \tau \eta ^A _s ds  \\
                  & - \int_0 ^t 2 \sqrt{ 2 \beta N_s } \hspace{1pt}
                                  \mathrm{Im} \Big( f \big( E_s , E_{s - \delta} , N_s \big) \Big) \tau \eta ^B _s ds \\
                  & + \int_0 ^t \frac{S'(N_s)}{\sqrt{2S (N_s)}} \hspace{1pt} h \big( E_s , N_s \big) \tau \eta ^C _s ds \; ,
\end{align*}
and
\begin{align*}
H^N _t = & \int_0 ^t 4 \beta \tau \Big( E^R_s \eta ^A _s + E^I_s \eta ^B _s \Big) ^2 ds  \hspace{95pt} \\
        & - \int_0 ^t 2 \sqrt{ \frac{ \beta S( N_s )}{ N_s }}
                                  \Big( E^R_s \tau \eta ^A _s + E^I_s \tau \eta ^B _s \Big) \eta ^C_s ds \\
        & - \int_0 ^t  4 \beta N_s \Big(  \tau \big( \eta ^A _s \big) ^2 
                           + \tau \big( \eta ^B _s \big) ^2 \Big) ds \\
        & - \int_0 ^t 2 \sqrt{ \frac{ \beta N_s }{S (N_s)}} S'(N_s)
                         \tau \eta ^C _s  \Big( E^R _s \eta ^A _s + E^I _s \eta ^B _s \Big) ds \\
        & + \int_0 ^t S'(N_s) \tau \big( \eta ^C _s \big) ^2 ds \; .
\end{align*}
Now, in view of (\ref{covariance function}), for all $i$ and all $s$, $\tau \eta ^i _s$ has mean zero and variance $\tau / 2$.
Hence, $\tau \eta ^i _s \rightarrow 0$ as the correlation time $\tau \rightarrow 0$.  Therefore, $U^E _t$ and $U^N _t$ converge to zero 
as $\tau \rightarrow 0$.  The terms $H^E _t$ and $H^N _t$ converge 
to the integrals of the additional drift terms in (\ref{limiting equation 1}) and (\ref{limiting equation 2}), respectively.  More precisely, as the correlation time 
$\tau \rightarrow 0$,
\begin{subequations}
\begin{flalign}
\label{additional field drift}
& - \int_0 ^t 2 \beta \Big( \tau \eta ^A _s + i \tau \eta ^B _s \Big) 
                    \Big( E^R _s \eta ^A _s + E^I _s \eta ^B _s \Big) ds \notag \hspace{60pt} \\
    & \hspace{120pt}  \longrightarrow \hspace{5pt} - \int_0^t \beta E_s ds \; , \\
  \label{additional field drift zero}
   &         \int_0 ^t 4 \beta \tau \Big( E^R_s \eta ^A _s + E^I_s \eta ^B _s \Big) ^2 ds
                                    \longrightarrow \int_0 ^t 2 \beta \left| E _s \right| ^2 ds \; , \\
   & - \int_0 ^t  4 \beta N_s \Big(  \tau \big( \eta ^A _s \big) ^2 
                           + \tau \big( \eta ^B _s \big) ^2 \Big) ds \notag \\
               & \hspace{115pt} \longrightarrow \hspace{5pt} - \int _0 ^t 4 \beta N_s ds \; , \\
  &  \int_0 ^t S'(N_s) \tau \big( \eta ^C _s \big) ^2 ds \hspace{4pt} \longrightarrow \hspace{4pt} \int _0 ^t \frac{1}{2} S'(N_s) ds \; , \\
  & \mathrm{and} \notag \\
  \label{last additional carrier drift}
  & \int_0 ^t \sqrt{ \frac{ \beta S( N_s ) }{ N_s }} 
                         \Big( \tau \eta ^A _s + i \tau \eta ^B _s \Big) \eta ^C_s ds \; , \notag \\
  & - \int_0 ^t 2 \sqrt{ \frac{ \beta S( N_s )}{ N_s }}
                                  \Big( E^R_s \tau \eta ^A _s + E^I_s \tau \eta ^B _s \Big) \eta ^C_s ds \; , \notag \\
  & - \int_0 ^t 2 \sqrt{ \frac{ \beta N_s }{S (N_s)}} S'(N_s)
                         \tau \eta ^C _s  \Big( E^R _s \eta ^A _s + E^I _s \eta ^B _s \Big) ds \notag \\
      & \hspace{165pt} \longrightarrow \hspace{3pt} 0 \; .
\end{flalign}
\end{subequations}
We will show (\ref{additional field drift}).  The limits (\ref{additional field drift zero}) -- (\ref{last additional carrier 
drift}) follow by the same general argument.  Let
\begin{equation*}
 I \equiv \int_0^t \bigg( 2 \Big( \tau \eta ^A _s + i \tau \eta ^B _s \Big) 
                    \Big( E^R _s \eta ^A _s + E^I _s \eta ^B _s \Big) - E_s \bigg) ds \; .
\end{equation*}
The limit (\ref{additional field drift}) is equivalent to $I \rightarrow 0$, which is a consequence of
\begin{subequations}
\begin{flalign}
\label{I first part}
 & I_1 \equiv \int_0^t \Big( 2 \tau \big( \eta ^A _s \big) ^2 E^R _s - E^R_s \Big) ds \hspace{3pt} \longrightarrow \hspace{3pt} 0 \; , \\
\label{I second part}
 & I_2 \equiv \int_0^t \Big( 2 i \tau \big( \eta ^B _s \big) ^2  E^I _s - i E^I_s \Big) ds \hspace{3pt} \longrightarrow \hspace{3pt} 0 \; , \\
  \mathrm{and} \hspace{5pt} & \hspace{225pt} \notag \\
 \label{I third part}
 & I_3 \equiv \int_0^t 2 \tau \eta ^A _s \eta ^B _s \Big( E^I _s + i E^R _s \Big) ds \hspace{3pt} \longrightarrow \hspace{3pt} 0  \; .
\end{flalign}
\end{subequations}
We show (\ref{I first part}); the limits (\ref{I second part}) and (\ref{I third part}) follow similarly.
We first define the new process $\tilde{\eta} ^A$ by $\tilde{\eta} ^A _s = \sqrt{\tau} \eta ^A _{\tau s}$.  
Then $\tilde{\eta} ^A$ solves the equation
\begin{equation}
\label{new OU process}
 d \tilde{\eta} ^A _s = - \tilde{ \eta } ^A _s ds + d \tilde{W} ^A _s \hspace{10pt}
\end{equation}
where $\tilde{W} ^A $ is the Wiener process defined by $\tilde{W} ^A _s = \tau ^{- 1/2} W ^A _{\tau s}$.  We express the integral $I_1$ in terms of this process:
\begin{equation*}
 I_1 = \int_0 ^t E^R _s \Big( 2 \big( \tilde{\eta} ^A _{s / \tau } \big) ^2 - 1 \Big) ds \; .
\end{equation*}
Let $\{ t_j : 0 \leq j \leq n \}$ be a partition of $[0,t]$ such that $t_{j + 1} - t_j = t / n = O ( \sqrt{\tau} )$.  We 
can then write
\begin{equation*}
 I_1 = \sum _{j = 0}^{n - 1} \int_{t_j}^{t_{j+1}} E^R _s \Big( 2 \big( \tilde{\eta} ^A _{s / \tau } \big) ^2 - 1 \Big) ds
\end{equation*}
or, equivalently,
\begin{equation*}
  I_1 = \tau \sum _{j = 0}^{n - 1} \int_{t_j / \tau} ^{t_{j+1} / \tau} E^R _{\tau r} \Big( 2 \big( \tilde{\eta} ^A _r \big) ^2 - 1 \Big) dr \; .
\end{equation*}
In each of the integrals in this sum the length of the interval of integration is $O ( \tau ^{- 1 / 2} )$.  Thus, since the slower variable $E^R$ evolves on a timescale of order 
one (in $\tau$), for $\tau$ sufficiently small we can approximate $I_1$ by
\begin{equation}
 \label{key integral split}
    \tau \sum _{j = 0}^{n - 1} E^R _{t_j} \int_{t_j / \tau} ^{t_{j+1} / \tau} \Big( 2 \big( \tilde{\eta} ^A _r \big) ^2 - 1 \Big) dr \; .
\end{equation}
We now estimate 
\begin{equation}
\label{key integral 2}
 \left| \int_{t_j / \tau} ^{t_{j+1} / \tau} \Big( 2 \big( \tilde{\eta} ^A _r \big) ^2 - 1 \Big) dr \right| \; .
\end{equation}
In view of (\ref{covariance function}), we have $E [ ( \tilde{\eta} ^A _r ) ^2 ] = 1 / 2$.  Therefore, for fixed $r$, the integrand in (\ref{key integral 2}) is a mean-zero random 
variable.  Thus, the integral in (\ref{key integral 2}) can be thought of as a sum of $O ( \tau ^{- 1 / 2} )$ identically 
distributed (by the stationarity of $\eta ^A$) mean-zero random variables.  Furthermore, equations (\ref{new OU process}) and (\ref{covariance function}) show that these random 
variables are weakly correlated, since 
the covariance function of 
the process $\tilde{\eta} ^A$ decays exponentially with an exponential decay constant equal to one.  Thus, in view of the law of 
large numbers, (\ref{key integral 2}) is on the order of $\tau ^{- 1 / 4}$ (this can also be shown by explicitly calculating the second moment of (\ref{key integral 2})).  Therefore,
since $n = O ( \tau ^{ - 1 / 2} )$, an upper bound for the order of 
(\ref{key integral split}) is $\tau ^{1 / 4}$ (this is, in fact, an overestimate since it does not take into 
account cancellations between positive and negative terms in the sum).  This estimate 
gives (\ref{I first part}).  The limits (\ref{I second part}), (\ref{I third part}), and (\ref{additional field drift zero}) -- (\ref{last additional carrier drift}) follow from the 
same general argument, using the fact that, for $i \neq j$, $E [ \eta ^i _r \eta ^j _r ] = 0$ by the independence of $\eta ^i$ and $\eta ^j$.  This completes 
the derivation of (\ref{limiting system}).

\section{Conclusion}

We have considered the problem of modeling spontaneous emission noise and carrier noise within the framework of stochastic differential equations.  
The stochastic differential equation model that we presented has corresponding diffusion coefficients that agree exactly with those
derived in the classic works of Henry \cite{Henry1982, Henry1983}.  While recent studies of the effects of the noise on the laser dynamics have neglected the correlations between the 
spontaneous emission noise and the carrier noise, our model 
describes the strong negative correlation between the intensity noise and the carrier noise.  More precisely, we have shown that the same Wiener processes that drive the spontaneous emission noise also drive 
the carrier noise.  This result could have important consequences for the dynamics.  For example, over any time interval the contribution to the carrier number of the part of the carrier noise 
that does not involve 
the electron-hole recombination rate $S$ is exactly the negative of the contribution of the spontaneous emission noise to the intensity of the electric field.  Clearly then, the inherent 
noise cannot be adequately described by previously used dynamical models in which the spontaneous emission noise and the carrier noise are uncorrelated.  In the presence of optical feedback 
in particular, there has been observed a disagreement between the
experimentally measured and the numerically calculated values of certain properties of the semiconductor laser system.  While the inclusion of noise in the dynamical model gives better agreement, previously 
used models do not satisfactorily reproduce for 
example features of the intensity distribution such as the narrow peak at large intensities \cite{Torcini}.

We have also shown that the nonzero correlation time of the physical noise affects the semiconductor laser dynamics in a systematic way.  The magnitude of these effects is independent of the size of the small 
correlation time of the inherent noise and thus does not vanish for exceedingly small correlation times.  The effects of the 
nonzero correlation time of the physical noise are captured by the additional drift terms in (\ref{limiting system}).  To be precise, these effects are the combined result of 
the dependence of the strength of the spontaneous emission noise on the carrier number, the dependence of the strength of the carrier noise on both the carrier number and the field, the correlations between the 
spontaneous emission noise and the carrier noise, and the nonzero correlation time of the noise.  
The system (\ref{limiting system}) reveals that the spontaneous emission noise and carrier noise affect the mean values of both the complex amplitude of the 
field and 
the carrier number. 
Comparing (\ref{limiting system}) with (\ref{LKequations}), we see that simply ignoring the noise terms in (\ref{LKequations}) does not yield the correct deterministic part of the system
because there is a contribution from the stochastic fluctuations.  This contribution is the result of correlations in the physical noise, which influence the mean dynamics of 
the semiconductor laser system.

\section*{Acknowledgments}
This work was supported by the Air Force Office of Scientific Research (grant number FA9550-15-1-0096).


\begin{thebibliography}{10}

\bibitem{Scully1966}
M.~Scully and W.~E. Lamb, Jr., ``Quantum theory of an optical maser,''
  \emph{Phys. Rev. Lett.}, vol. 16(19), pp. 853--855, 1966.

\bibitem{Scully1967}
M.~Scully and W.~E. Lamb, Jr., ``Quantum theory of an optical maser. i. general theory,'' \emph{Phys.
  Rev.}, vol. 159(2), pp. 208--226, 1967.

\bibitem{Lax1966}
M.~Lax, ``Quantum noise. {IV}. {Quantum} theory of noise sources,'' \emph{Phys.
  Rev.}, vol. 145(1), pp. 110--129, 1966.

\bibitem{Lax}
M.~Lax, ``Quantum noise {VII}: The rate equations and amplitude noise in
  lasers,'' \emph{IEEE J. Quantum Electron.}, vol. QE-3(2), pp. 37--46, 1967.

\bibitem{Henry1982}
C.~H. Henry, ``Theory of the linewidth of semiconductor lasers,'' \emph{IEEE J.
  Quantum Electron.}, vol. 18(2), pp. 259--264, 1982.

\bibitem{Henry1983}
C.~H. Henry, ``Theory of the phase noise and power spectrum of a single mode
  injection laser,'' \emph{IEEE J. Quantum Electron.}, vol. 19(9), pp.
  1391--1397, 1983.

\bibitem{Chow}
W.~Chow, S.~Koch, and M.~Sargent~III, \emph{Semiconductor-Laser Physics}.\hskip
  1em plus 0.5em minus 0.4em\relax Berlin: Springer-Verlag, 1994.

\bibitem{Carmichael}
H.~J. Carmichael, \emph{Statistical Methods in Quantum Optics 1: Master
  Equations and Fokker-Planck Equations}.\hskip 1em plus 0.5em minus
  0.4em\relax Berlin: Springer-Verlag, 1999.

\bibitem{Meystre}
P.~Meystre and M.~Sargent~III, \emph{Elements of Quantum Optics}.\hskip 1em
  plus 0.5em minus 0.4em\relax Berlin: Springer-Verlag, 1999.

\bibitem{Mulet}
J.~Mulet and C.~R. Mirasso, ``Numerical statistics of power dropouts based on
  the {Lang-Kobayashi} model,'' \emph{Phys. Rev. E}, vol. 59(5), pp.
  5400--5405, 1999.

\bibitem{Giacomelli}
G.~Giacomelli, M.~Giudici, S.~Balle, and J.~R. Tredicce, ``Experimental
  evidence of coherence resonance in an optical system,'' \emph{Phys. Rev.
  Lett.}, vol. 84(15), pp. 3298--3301, 2000.

\bibitem{Buldu2001}
J.~M. Buld{\'u}, J.~Garc{\'i}a-Ojalvo, C.~R. Mirasso, M.~C. Torrent, and J.~M.
  Sancho, ``Effect of external noise correlation in optical coherence
  resonance,'' \emph{Phys. Rev. E}, vol. 64(5), pp. 051\,109--1--4, 2001.

\bibitem{Buldu2002}
J.~M. Buld{\'u}, J.~Garc{\'i}a-Ojalvo, C.~R. Mirasso, and M.~C. Torrent,
  ``Stochastic entrainment of optical power dropouts,'' \emph{Phys. Rev. E},
  vol.~66, pp. 021\,106--1--5, 2002.

\bibitem{Buldu2004}
J.~M. Buld{\'u}, J.~Garc{\'i}a-Ojalvo, and M.~C. Torrent, ``Delay-induced
  resonances in an optical system with feedback,'' \emph{Phys. Rev. E},
  vol.~69, pp. 046\,207--1--5, 2004.

\bibitem{Marino}
F.~Marino, M.~Giudici, S.~Barland, and S.~Balle, ``Experimental evidence of
  stochastic resonance in an excitable optical system,'' \emph{Phys. Rev.
  Lett.}, vol. 88(4), pp. 040\,601--1--4, 2002.

\bibitem{Yousefi}
M.~Yousefi, D.~Lenstra, and G.~Vemuri, ``Carrier inversion noise has important
  influence on the dynamics of a semiconductor laser,'' \emph{IEEE J. Sel. Top.
  Quantum Electron.}, vol. 10(5), pp. 955--960, 2004.

\bibitem{Torcini}
A.~Torcini, S.~Barland, G.~Giacomelli, and F.~Marin, ``Low-frequency
  fluctuations in vertical cavity lasers: Experiments versus {Lang-Kobayashi}
  dynamics,'' \emph{Phys. Rev. A}, vol. 74(6), pp. 063\,801--1--13, 2006.

\bibitem{Soriano2011}
M.~C. Soriano, T.~Berkvens, G.~Van~der Sande, G.~Verschaffelt, J.~Danckaert,
  and I.~Fischer, ``Interplay of current noise and delayed optical feedback on
  the dynamics of semiconductor lasers,'' \emph{IEEE J. Quantum Electron.},
  vol. 47(3), pp. 368--374, 2011.

\bibitem{VanderSande}
G.~Van~der Sande, M.~C. Soriano, M.~Yousefi, M.~Peeters, J.~Danckaert, and
  G.~Verschaffelt, ``Influence of current noise on the relaxation oscillation
  dynamics of semiconductor lasers,'' \emph{Appl. Phys. Lett.}, vol.~88, pp.
  071\,107--1--3, 2006.

\bibitem{Yamada}
M.~Yamada, \emph{Theory of Semiconductor Lasers}, ser. Springer Series in
  Optical Sciences.\hskip 1em plus 0.5em minus 0.4em\relax Tokyo: Springer,
  2014, vol. 185.

\bibitem{Kane}
D.~M. Kane and K.~A. Shore, Eds., \emph{Unlocking Dynamical Diversity: Optical
  Feedback Effects on Semiconductor Lasers}.\hskip 1em plus 0.5em minus
  0.4em\relax Chichester, England: Wiley, 2005.

\bibitem{Gardiner}
C.~W. Gardiner, \emph{Handbook of Stochastic Methods}.\hskip 1em plus 0.5em
  minus 0.4em\relax Berlin: Springer-Verlag, 1985.

\bibitem{Oksendal}
B.~{\O}ksendal, \emph{Stochastic Differential Equations: An Introduction with
  Applications}.\hskip 1em plus 0.5em minus 0.4em\relax Berlin:
  Springer-Verlag, 1998.

\bibitem{Barland}
S.~Barland, P.~Spinicelli, G.~Giacomelli, and F.~Marin, ``Measurement of the
  working parameters of an air-post vertical-cavity surface-emitting laser,''
  \emph{IEEE J. Quantum Electron.}, vol. 41(10), pp. 1235--1243, 2005.

\bibitem{Wong1965}
E.~Wong and M.~Zakai, ``On the relation between ordinary and stochastic
  differential equations,'' \emph{Int. J. Eng. Sci.}, vol.~3, pp. 213--229,
  1965.

\bibitem{Nelson}
E.~Nelson, \emph{Dynamical Theories of Brownian Motion}.\hskip 1em plus 0.5em
  minus 0.4em\relax Princeton: Princeton University Press, 1967.

\end{thebibliography}
\end{document}